\documentclass[preprint,aps]{revtex4}

\usepackage{stmaryrd}
\usepackage{amstext}
\usepackage{amsfonts}
\usepackage{amsmath}
\usepackage{amssymb}
\usepackage{epsfig}
\usepackage{graphicx}%
\ifx\pdfoutput\relax\let\pdfoutput=\undefined\fi
\newcount\msipdfoutput
\ifx\pdfoutput\undefined\else
\ifcase\pdfoutput\else
\ifx\paperwidth\undefined\else
\ifdim\paperheight=0pt\relax\else\pdfpageheight\paperheight\fi
\ifdim\paperwidth=0pt\relax\else\pdfpagewidth\paperwidth\fi
\fi\fi\fi
\begin{document}

\title{Entanglement generation and manipulation in the Hong-Ou-Mandel experiment: A hidden scenario beyond two-photon interference}

\author{Li-Kai Yang$\footnote{email:xsylk@mail.ustc.edu.cn}$}
\affiliation{Institute of Quantum Science and Engineering, Texas A$\&$M University, Texas 77843}

\author{Han Cai}
\affiliation{Institute of Quantum Science and Engineering, Texas A$\&$M University, Texas 77843}

\author{Tao Peng}
\affiliation{Institute of Quantum Science and Engineering, Texas A$\&$M University, Texas 77843}

\author{Da-Wei Wang}
\affiliation{Institute of Quantum Science and Engineering, Texas A$\&$M University, Texas 77843}

\begin{abstract}
Hong-Ou-Mandel (HOM) effect was long believed to be a two-photon interference phenomenon. It describes the fact that two indistinguishable photons mixed at a beam splitter will bunch together to one of the two output modes. Considering the two single-photon emitters such as trapped ions, we explore a hidden scenario of the HOM effect, where entanglement can be generated between the two ions when a single photon is detected by one of the detectors. A second photon emitted by the entangled photon sources will be subsequently detected by the same detector. However, we can also control the fate of the second photon by manipulating the entangled state. Instead of two-photon interference, phase of the entangled state is responsible for photon's path in our proposal. Toward a feasible experimental realization, we conduct a quantum jump simulation on the system to show its robustness against experimental errors.
\end{abstract}

\maketitle

\section{Introduction}

In 1987, Hong, Ou and Mandel did an experiment to measure the time interval between two photons generated from spontaneous parametric down conversion (SPDC) \cite{Hong}. They show that when two indistinguishable photons are mixed at a beam splitter, they will bunch with each other into either one of the two outputs. This phenomenon of two-photon interference is later known as the Hong-Ou-Mandel (HOM) effect. The HOM-type interference have also been found in electrons \cite{Bocquillon,Dubois}, atoms \cite{Kaufman,Lopes} and phonons \cite{Toyoda}. The development of ion traps \cite{Paul} and other technologies made the observation and manipulation of single particle possible. As a result, different versions of HOM experiments have been done by using quantum dots \cite{Santori,Patel}, atoms in cavity quantum electrodynamics \cite{Legero} and nearby trapped neutral atoms \cite{Beugnon} as photon sources. It has also been found that entanglement can be generated between photon sources (e.g. atoms and ions). Different protocols of entanglement generation using such effect have been proposed \cite{Feng,Cabrillo,Yang,Moehring,Hong2,Simon,Zhao}. In addition, HOM experiment has been used to realize remote quantum communication \cite{Duan,Bose} and assess photon entanglement \cite{Barbieri}.

In this paper, we introduce a protocol for generating entanglement based on HOM experiment \cite{Browne} and prove its robustness against experimental imperfections. We will show that two photons after the beam splitter can be detected one after another by the same detector and in the time interval between the two detections, the light sources are entangled. This is different from the photon interference picture of the original HOM experiment, where no entanglement between the light sources are involved. The entanglement between the two light sources can be further manipulated such that the path of the second photon can be controlled. Starting from the same setup as the original HOM experiment, we reconstruct its result by introducing entangled light sources (trapped ions in our case). This gives us a new insight into HOM effect that it can be interpreted by entanglement rather than photon interference.

The article is organized as following. We show our general idea in Sec. 2. In Sec. 2 A, we use the spontaneous emission of two-level atoms as photon source to repeat the original HOM experiment. Toward a feasible experimental realization, we introduce an improved setup with Raman transitions in Sec. 2 B. We show that in such a system, the entangled state induced by single-photon detection can be manipulated to achieve a path redistribution of the second photon. In Sec. 3, we conduct a quantum jump simulation of our system to show its robustness against various kinds of experimental noises.

\section{The basic theory}

\subsection{Simplified model using spontaneous emission}
\begin{figure*}
\includegraphics[width=0.95\linewidth]{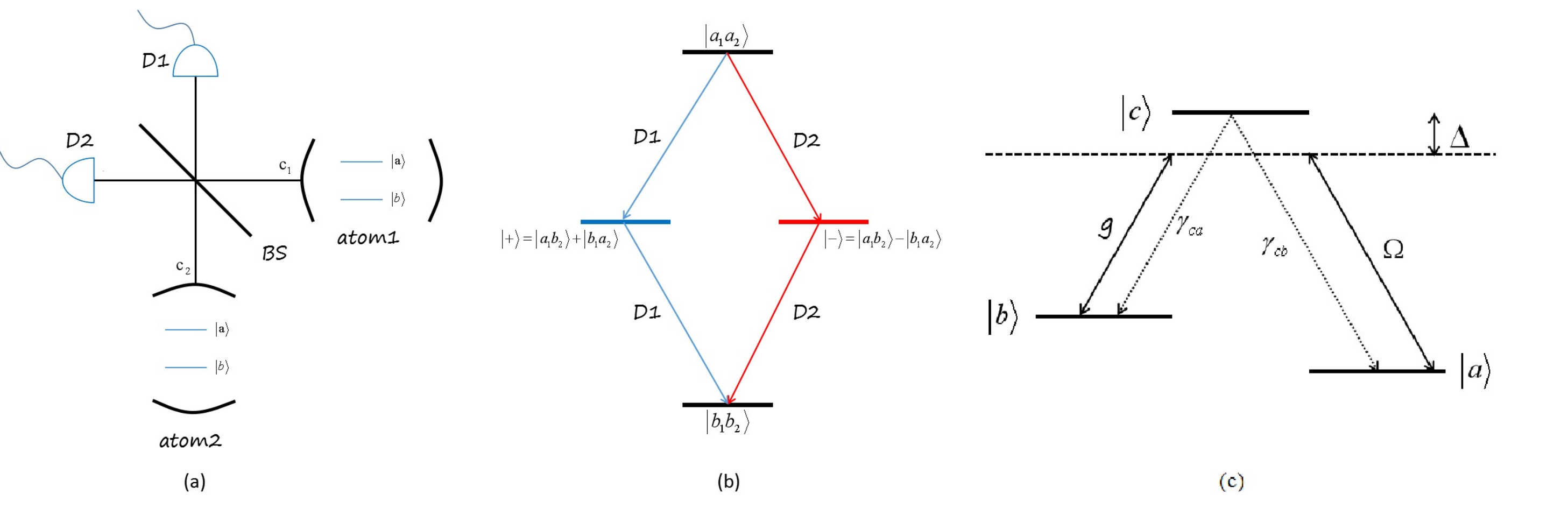}
\centering
\caption{(a) A toy model of the proposed scheme. The spontaneously emitted photons from two identical two-level atoms mix at a $50/50$ beam splitter. (b) Basic process of our scheme. When a photon is detected by $D1$ (or $D2$), the two-atom state will be projected into an entangled state $|+\rangle$ (or $|-\rangle$), from which the second photon is subsequently emitted and observed by the same detector. (c) Internal physical levels of Raman-like transition system.}
\label{f1}
\end{figure*}

In order to demonstrate the physical picture of our scheme, in the following we briefly introduce a toy model. The experimental setup of this model is sketched in Fig. \ref{f1}(a). Two identical two-level atoms with excited state $|a\rangle$ and ground state $|b\rangle$ are trapped in two separate cavities. Photons emitted from spontaneous decay leak from the cavity and mix at a $50/50$ beam splitter. We assume that the coupling between the atoms and the cavities satisfy the bad cavity limit and only an enhanced decay rate of the atoms is introduced by the cavities. The photons from the outputs of the beam splitter can be observed by two detectors.

According to Weisskopf-Wigner theory, the time evolution of the wave function is
\begin{equation}
\begin{split}
&|\psi(t)\rangle=e^{-\Gamma{t}}|a_1a_2\rangle|0\rangle+\sqrt{\frac{\Gamma}{2\pi}}\int\text{d}\nu_1 [\frac{1-e^{i(\omega-\nu_1)t-\Gamma{t}/2}}{(\nu_1-\omega)+i\Gamma/2}]e^{-\Gamma{t}/2}|b_1a_2\rangle|1_{\nu_1}0\rangle \\
&+\sqrt{\frac{\Gamma}{2\pi}}\int\text{d}\nu_2 [\frac{1-e^{i(\omega-\nu_2)t-\Gamma{t}/2}}{(\nu_2-\omega)+i\Gamma/2}]e^{-\Gamma{t}/2}|a_1b_2\rangle|01_{\nu_2}\rangle\\
&+{\frac{\Gamma}{2\pi}}\int\text{d}\nu_1\text{d}\nu_2[\frac{1-e^{i(\omega-\nu_1)t-\Gamma{t}/2}}{(\nu_1-\omega)+i\Gamma/2}][\frac{1-e^{i(\omega-\nu_2)t-\Gamma{t}/2}}{(\nu_2-\omega)+i\Gamma/2}]|b_1b_2\rangle|1_{\nu_1}1_{\nu_2}\rangle,
\end{split}
\label{wf}
\end{equation}
where $\Gamma$ is the cavity-enhanced decay rate. $|b_1a_2\rangle$ means the first atom is in the ground state $|b\rangle$ and the second atom is in the excited state $|a\rangle$. $|1_{\nu_1(\nu_2)}\rangle$ means the single photon state emitted by the first (second) atom with frequency $\nu_1 (\nu_2)$. We assume the optical path length from the atoms to the detectors is $L$. If $L/c\gg 1/\Gamma$, i.e., when the detectors detect the photons, the two atoms have decayed. The last term dominates the wavefunction. Thus we have,
\begin{equation}
|\psi(t)\rangle_{HOM}\approx{\frac{\Gamma}{2\pi}}\int\text{d}\nu_1\text{d}\nu_2[\frac{1-e^{i(\omega-\nu_1)t-\Gamma{t}/2}}{(\nu_1-\omega)+i\Gamma/2}][\frac{1-e^{i(\omega-\nu_2)t-\Gamma{t}/2}}{(\nu_2-\omega)+i\Gamma/2}]|b_1b_2\rangle|1_{\nu_1}1_{\nu_2}\rangle \varpropto|b_1b_2\rangle|1_{\gamma_1}1_{\gamma_2}\rangle,
\end{equation}
where $|1_{\gamma_j}\rangle=\sqrt{t}\int_{\omega-1/2t}^{\omega+1/2t}\text{d}\nu_j|1_{\nu_j}\rangle$ is a single photon state with a frequency width mainly determined by the detection time $t$ (since it is much shorter than the life time of the atoms). Consequently, after the beam splitter, the single photon states are transformed to the detector modes,
\begin{equation}
\begin{split}
&|1_{\gamma_1}\rangle=(|1_{\gamma_a}\rangle+|1_{\gamma_b}\rangle)/\sqrt{2},\\
&|1_{\gamma_2}\rangle=(|1_{\gamma_a}\rangle-|1_{\gamma_b}\rangle)/\sqrt{2},
\end{split}
\label{bs}
\end{equation}
where $|1_{\gamma_a}\rangle$ and $|1_{\gamma_b}\rangle$ are single photon states that can be detected by detectors $D_1$ and $D_2$, respectively. The conventional HOM effect is shown by
\begin{equation}
|1_{\gamma_1}1_{\gamma_2}\rangle=\frac{|2_{\gamma_a}0_{\gamma_b}\rangle-|0_{\gamma_a}2_{\gamma_b}\rangle}{\sqrt2}.
\end{equation}
Up to here we demonstrate that the HOM experiment can be repeated via photons generated from spontaneous decay. However, if $L/c\ll 1/\Gamma$, the detectors can detect a photon when probably only one photon is emitted. To the first order of $t\ll1/\Gamma$, the last term in Eq. (\ref{wf}) can be neglected and the second and the third terms in Eq. (\ref{wf}) are
\begin{equation}
\begin{split}
&-it\sqrt{\frac{\Gamma}{2\pi}}\int\limits_{\omega-1/2t}^{\omega+1/2t}\text{d}\nu_1|b_1a_2\rangle|1_{\nu_1}0\rangle -it\sqrt{\frac{\Gamma}{2\pi}}\int\limits_{\omega-1/2t}^{\omega+1/2t}\text{d}\nu_2|a_1b_2\rangle|01_{\nu_2}\rangle\\
&=-i\sqrt{\frac{\Gamma t}{2\pi}}(|b_1a_2\rangle|1_{\gamma_1}0\rangle+|a_1b_2\rangle|01_{\gamma_2}\rangle).
\end{split}
\label{sp}
\end{equation}
Substituting Eq. (\ref{bs}) in Eq. (\ref{sp}), we obtain,
\begin{equation}
\begin{split}
-i\sqrt{\frac{\Gamma t}{4\pi}}[(|b_1a_2\rangle+|a_1b_2\rangle)|1_{\gamma_a}\rangle+(|b_1a_2\rangle-|a_1b_2\rangle)|1_{\gamma_b}\rangle].
\end{split}
\end{equation}
When a photon is observed by one of the detectors, the atomic state will be projected into one of the two maximally entangled states,
\begin{equation}
|\pm\rangle=\frac{1}{\sqrt2}(|b_1a_2\rangle\pm|a_1b_2\rangle).
\end{equation}
Starting from these entangled states, the two atoms will continue to emit another photon that can only be observed by the clicked detector due to the phase correlation established by the detection of the first photon. Eventually, we will again get the final state $\frac{|2_{\gamma_a}0_{\gamma_b}\rangle-|0_{\gamma_a}2_{\gamma_b}\rangle}{\sqrt2}$. The general evolution of this system is shown in Fig. \ref{f1}(b). In this scenario, we do not have two-photon interference. However, only one of the detectors will click and click twice, which is consistent with the traditional HOM experiments.

Although this system gives us a clear picture of the scheme, it may not be feasible in experiments. The spontaneous decay of an atom cannot be controlled easily, which limits our ability to control the entangled states. Also, we cannot guarantee that the photons emitted from the spontaneous decay will go into the beam splitter, which sharply lowers the probability of success in this scheme. In order to resolve these problems, we proposed a modified scheme with Raman transitions.

\subsection{Controllable system with Raman transitions}

To introduce the Raman-like transition system, we simply replace the two-level atoms in Fig. \ref{f1}(a) by trapped ions. The energy levels of ions are shown in Fig. \ref{f1}(c). They have three energy levels where the two lower levels $|a\rangle$ and $|b\rangle$ are coupled with the upper level $|3\rangle$ by a cavity field and a classical light field. A same large detuning $\Delta$ is introduced on both transitions. The $|b\rangle\leftrightarrow|c\rangle$ transition is coupled to the cavity mode while $|a\rangle\leftrightarrow|c\rangle$ transition is driven by the light field. If the initial state of an ion is $|a\rangle$, it is clear that a photon will be released to the cavity when the transition $|a\rangle\rightarrow|b\rangle$ happens.

To describe the evolution of such a system, we use a master equation approach. The Hamiltonian of the ion-cavity system can be written as
\begin{small}
\begin{equation}
H=\sum_{i=1,2}(\Delta|c\rangle_{ii}\langle{c}|+g|c\rangle_{ii}\langle{b}|c_i+g|b\rangle_{ii}\langle{c}|c_i^{\dagger}+\Omega|c\rangle_{ii}\langle{a}|+\Omega|a\rangle_{ii}\langle{c}|),
\end{equation}
\end{small}
where $\Omega$ and $g$ denote the coupling of the driving light field and the cavity mode, respectively, while $c_i$ is the annihilation operator for cavity mode. The subscript $i=1,2$ represents two individual cavities. Here we choose the interaction picture and set $\hbar=1$. To describe the dynamics of the system, we introduce an effective Hamiltonian used in quantum dissipation system \cite{Browne},
\begin{equation}
H_{eff}=H-i\kappa\sum_{i=1,2}c_i^{\dagger}c_i-i(\gamma_{ca}+\gamma_{cb})\sum_{i=1,2}|c\rangle_{ii}\langle{c}|.
\end{equation}
The upper level $|c\rangle$ can decay to the two lower levels with decay rates $\gamma_{ca}$ and $\gamma_{cb}$, respectively. The decay rate of the cavities is $2\kappa$. This effective Hamiltonian is used to describe the evolution of the system when neither a spontaneous emission nor a cavity decay occurs. Under optimal conditions, we assume that $\gamma_{ca}=\gamma_{cb}=0$ and the detuning $\Delta$ is large enough, then $H_{eff}$ can be reduced to the form
\begin{small}
\begin{equation}
H'_{eff}=\sum_{i=1,2}[\frac{g\Omega}{\Delta}(|a\rangle_{ii}\langle{b}|c_i+|b\rangle_{ii}\langle{a}|c_i^{\dagger})+\frac{g^2}{\Delta}|b\rangle_{ii}\langle{b}|+\frac{\Omega^2}{\Delta}|a\rangle_{ii}\langle{a}|-i{\kappa}c_i^{\dagger}c_i]
\end{equation}
\end{small}
When no cavity decay happens, the system evolves under $H'_{eff}$. Consider the initial state $|\psi(0)\rangle=|aa\rangle|00\rangle$, then when the evolution time $t$ is short enough to obey the relationship $\frac{g{\Omega}t}{\Delta}\ll1$, the state evolves as
\begin{equation}
\begin{split}
|\psi(t)\rangle=& e^{-iH'_{eff}t}|\psi(0)\rangle\approx(1-iH'_{eff}t)|\psi(0)\rangle \\
& =(1-i\frac{2\Omega^2t}{\Delta})|aa\rangle|11\rangle-i\frac{g\Omega{t}}{\Delta}(|ba\rangle|10\rangle+|ab\rangle|01\rangle),
\end{split}
\end{equation}
where $|1\rangle$ and $|0\rangle$ denote the photon state in cavity. The annihilation operators $d_1$ and $d_2$ of the output modes that can be detected by the photon detectors are related to the cavity modes $c_1$ and $c_2$ by
\begin{align}
&d_1=\frac{1}{\sqrt2}(c_1+c_2), \label{bs1}\\
&d_2=\frac{1}{\sqrt2}(c_1-c_2). \label{bs2}
\end{align}
If a detector clicks, let's say detector $d_1$ clicks, the state is projected into
\begin{equation}
|\psi_1\rangle=d_1|\psi(t)\rangle=\frac{1}{\sqrt2}(|ba\rangle+|ab\rangle)|00\rangle,
\end{equation}
with a detection rate $R\approx4\kappa(\frac{g\Omega}{\Delta\kappa})^2$. The detection of a single photon results in an entangled state of the two atoms.

In order to manipulate the entangled state, we apply a laser field to one of the ions. The ac Stark shift \cite{Aulter} due to the laser field can shift the energy level of the ion, which results in an additional energy difference between $|a\rangle$ and $|b\rangle$. Energy shift introduced by the laser field are highly sufficient and can be measured precisely \cite{Sherman}. By denoting the energy shift as $E$ and the duration of the applied laser field as $\tau$, the resulted entangled state is
\begin{equation}
|\psi_e\rangle=\frac{1}{\sqrt2}(|ba\rangle+e^{iE\tau}|ab\rangle)|00\rangle.
\end{equation}
After that we eliminate the laser field and turn on the driven field to drive the system, which evolves as
\begin{equation}
\begin{split}
|\psi_e(t)\rangle=&e^{-iH'_{eff}t}|\psi_e\rangle\approx(1-iH'_{eff}t)|\psi_e\rangle \\
&=\frac{1}{\sqrt2}(1-i\frac{g^2}{\Delta}t-i\frac{\Omega^2}{\Delta}t)(|ba\rangle+e^{iE\tau}|ab\rangle)|00\rangle+\frac{g\Omega{t}}{\sqrt2\Delta}|bb\rangle(|10\rangle+e^{iE\tau}|01\rangle).
\end{split}
\end{equation}
Thus, when the second photon is detected, the probability $P_1$ and $P_2$ at which detector $d_1$ or $d_2$ clicks obey the relation
\begin{equation}
P_1:P_2=\langle\psi_e(t)|d_1^{\dagger}d_1|\psi_e(t)\rangle:\langle\psi_e(t)|d_2^{\dagger}d_2|\psi_e(t)\rangle=(1+cosE\tau):(1-cosE\tau).
\label{redistribution}
\end{equation}
Notice that when $E\tau=0$, i.e., if we do not apply the laser field to manipulate the entangled state, the original HOM effect is achieved. However, when $E\tau=\pi$ it turns out that $P_1=0$, i.e. the second photon must be observed by detector $d_2$.

The process is similar if the first photon is detected by $d_2$. With the above calculation, we demonstrate the process of generating entangled states and the manipulation of the entangled states, which controls the fate of the second photon.

\section{quantum jump simulation}
Besides analytical calculation, it is necessary to prove the robustness and versatility of our scheme in experiments, where errors and imperfection are inevitable. We use a quantum jump simulation based on Monte Carlo method to simulate the experiment. Several kinds of experimental errors are included in the simulation to test the robustness of the above protocol. We briefly introduce the procedure of the simulation in the fo
owing \cite{Plenio}:
\begin{enumerate}
\item Set an initial state $|\psi(0)\rangle$ and a minimal time interval $\delta{t}$. The process will be divided into discrete period of time $\delta{t}$.
\item Determine the current probability of a quantum jump $P$, which is induced by a cavity decay or a spontaneous decay. For example, the probability that detector $d_1$ clicks at current time is $P_1=2\kappa\delta{t}\langle\psi(t)|d_1^{\dagger}d_1|\psi(t)\rangle$.
\item Generate a random number $r$ between $0$ and $1$ and compare with $P$.
\item If $r<P$ there is a quantum jump (decay). Since there might be multiple possibilities of quantum jump (e.g. detector $d_1$ clicks or spontaneous decay $|c\rangle\rightarrow|b\rangle$, etc.), we need to divide $P$ into various kinds of quantum jumps proportional to their probabilities. If one of the quantum jumps happens, for example, detector $d_1$ clicks, then the system jumps to the renormalized form
\begin{equation}
|\psi\rangle\rightarrow\frac{d_1|\psi\rangle}{\sqrt{\langle\psi|d_1^{\dagger}d_1|\psi\rangle}}.
\end{equation}
\item If $r>P$ no jump takes place, thus the system evolves under the effective Hamiltonian in a renormalized way, which gives
\begin{equation}
|\psi\rangle\rightarrow\frac{e^{-iH_{eff}\delta{t}}|\psi\rangle}{\sqrt{1-P}}.
\end{equation}
\item Repeat the above process for time interval $\delta{t}$ until a photon is detected or arriving at the maximal waiting time.
\end{enumerate}

It is clear that our scheme can be divided into two parts, namely, entanglement generation and manipulation, which can be marked by two single-photon detections. Therefore, in the following we demonstrate the simulation results of these two parts respectively.

\subsection{Entanglement generation}
The most common errors in the procedure of generating entanglement include detection efficiency, beam splitter imperfection and spontaneous decay. Hence, we consider these three errors in our simulation. We choose the following parameters: $\Omega=g$, $\kappa=10g$ and $\Delta=20g$. The theoretical probability for successfully getting entangled state is $p=RT=4\kappa{T}(g\Omega/\Delta\kappa)^2$ where $T$ is the waiting time. Choosing $T=100/g$, we get $p=0.1$. For each run of simulation, the process is conducted by $10^5$ times and the fidelity of entangled state ($F$) together with the success probability ($p$) being calculated.

\begin{figure}
\includegraphics[width=0.95\linewidth]{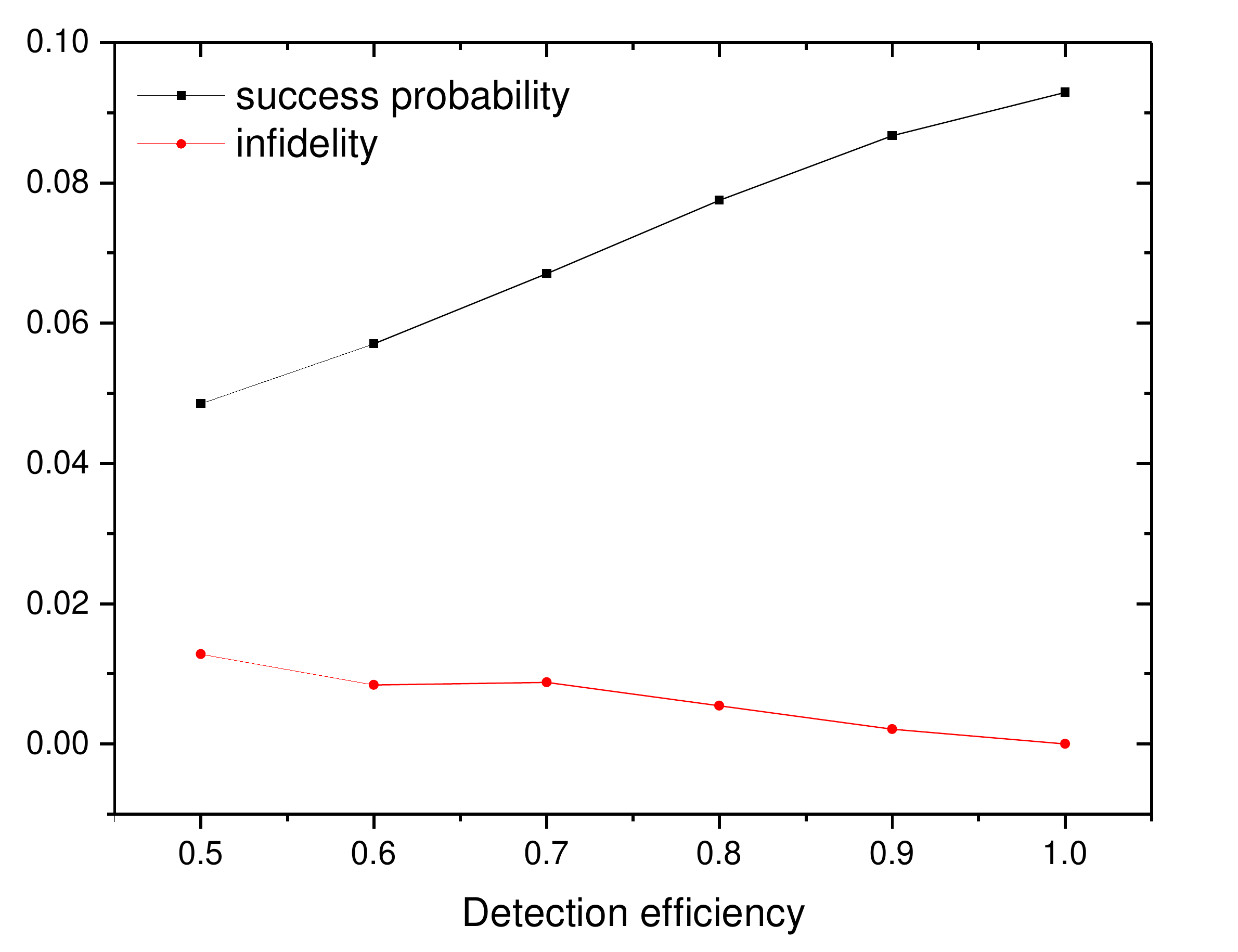}
\centering
\caption{Simulation of entanglement generation under different detection efficiency $\eta$, the simulation is conducted individually for $10^5$ times and the expectation value of success probability $p$ and infidelity $1-F$ are shown.}
\label{f2}
\end{figure}

$\emph{Detection efficiency:}$
In theoretical computation we assume that photons generated by Raman-like transition will surely be observed by one of the detectors. However, this assumption may not hold true in practical experiments. The decreased efficiency of detecting photons can be caused by several reasons. Namely, non-unitary detector efficiency and coupling efficiency between the ion and the cavity. By denoting the detection efficiency as $\eta$, it is quite straightforward that the success probability $p'$ under such condition will be $p'=\eta{p}=4\eta\kappa{T}(g\Omega/\Delta\kappa)^2$. The results of simulation with different $\eta$ are shown in Fig. \ref{f2}. According to the figure, the linear relation between $p$ and $\eta$ is well satisfied. Also, we can get an entangled state with $F\approx0.987$ even if $\eta$ reduced to $0.5$, which imply great robustness against imperfect detection efficiency.

\begin{figure}
\includegraphics[width=0.95\linewidth]{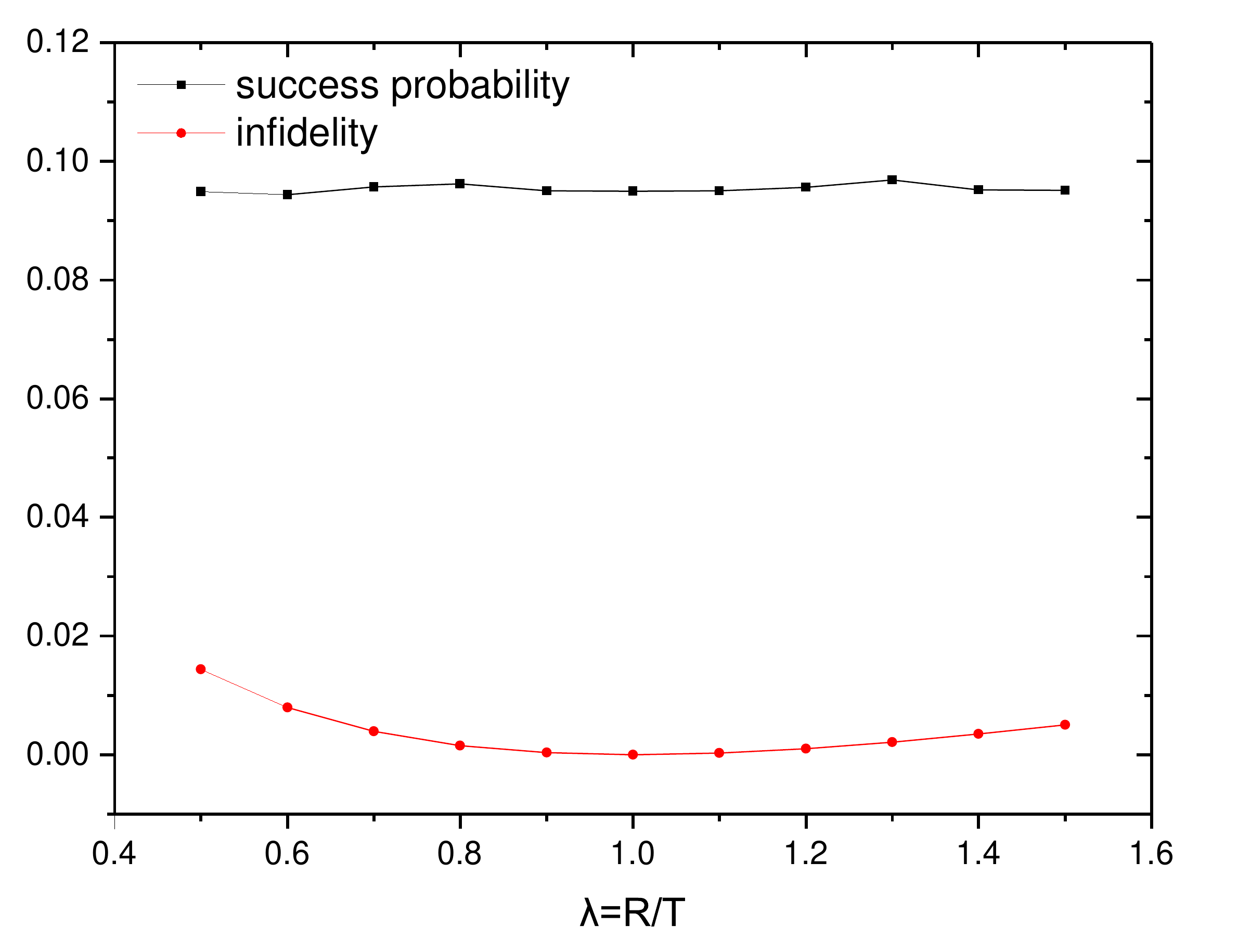}
\centering
\caption{Simulation of entanglement generation with beam splitter imperfection, i.e. different ratios $\lambda=R/T$.}
\label{f3}
\end{figure}

$\emph{Beam splitter imperfection:}$
In above discussion we consider an exact 50/50 beam splitter, which might be unrealistic in practical experiments. Instead, the ratio $\lambda=R/T$, where $R$ and $T$ represent the reflection and transition rate of beam splitter respectively, may not equal to one. This results in a changed form of Eq. (\ref{bs1}) and (\ref{bs2}) as
\begin{align}
& d_1=\sqrt{R}c_1+\sqrt{T}c_2, \\
& d_2=\sqrt{T}c_1-\sqrt{R}c_2.
\end{align}
By applying above relations to the simulation, we choose $\lambda$ ranging from $0.5$ to $1.5$ and the results are shown in Fig. \ref{f3}. From the figure we can conclude that the imperfection of beam splitter have little influence on either fidelity or success probability of the scheme. The entangled state can still reach a great fidelity of $F\approx0.99$ when $\lambda$ is as large as $1.5$, which is definitely an important feature of practicability.

\begin{figure}
\includegraphics[width=0.95\linewidth]{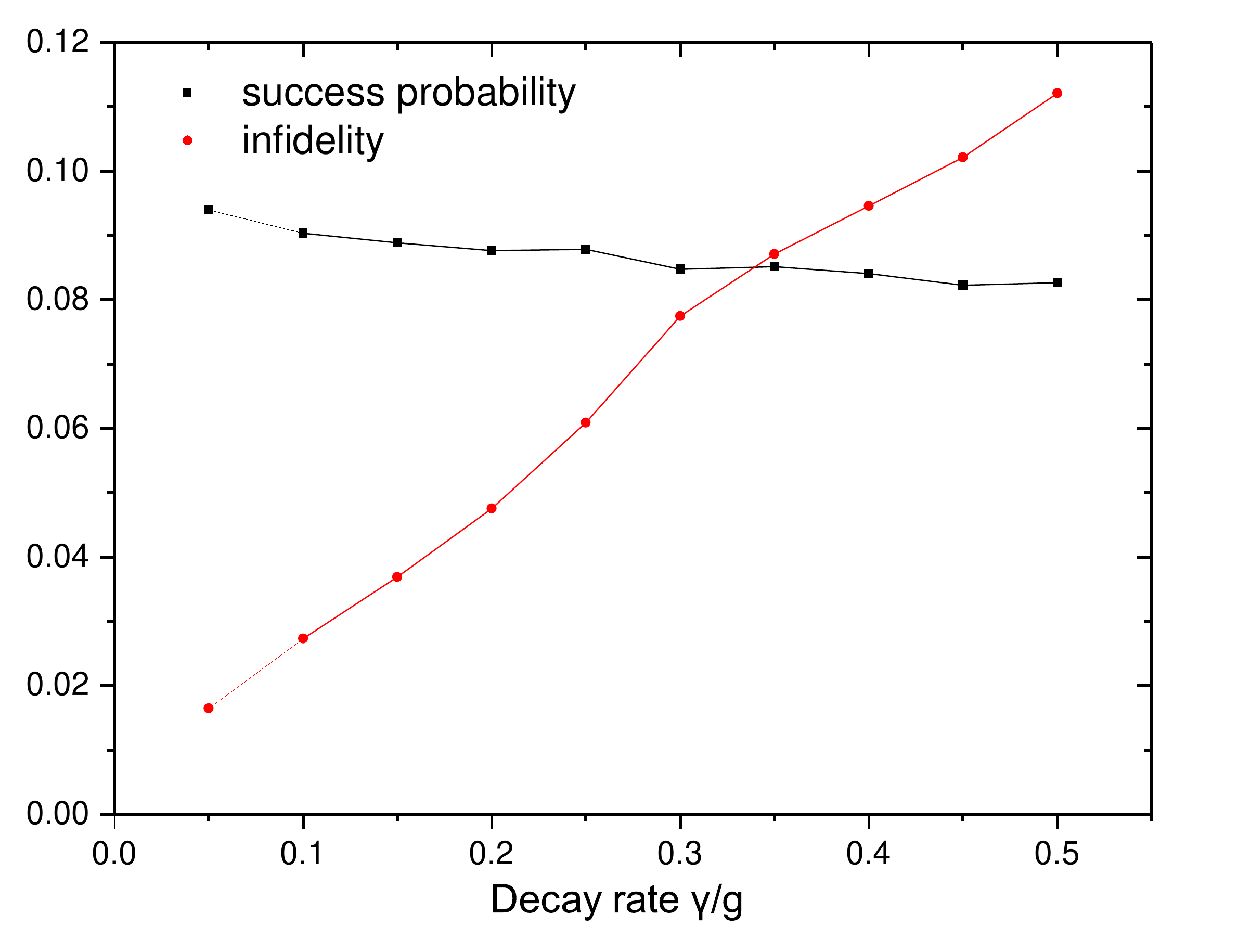}
\centering
\caption{Simulation of entanglement generation with different spontaneous decay rates $\gamma=\gamma_{cb}=\gamma_{ca}$.}
\label{f4}
\end{figure}

$\emph{Spontaneous decay:}$
Although Raman-like transition system provides us with the ability to control and adjust the trapped ions, spontaneous decay is still unavoidable in such system. As discussed in Sec. 2, when considering the effect of spontaneous decay, we must use $H_{eff}$ as evolution Hamiltonian instead of $H'_{eff}$. Moreover, in the process of simulation, multiple choices of quantum jump need to be included. For instance, if the spontaneous decay $|c\rangle\rightarrow|b\rangle$ takes place, no photon will be coupled into the cavity mode, hence results no possibility of photon detection. This will reduce the success probability of our scheme. For simplicity, in the simulation we set $\gamma_{ca}=\gamma_{cb}=\gamma$ and choose different $\gamma$ to verify its impact. Here we consider $\gamma$ ranging from $0.05g$ to $0.5g$ and the results are shown in Fig. \ref{f4}. We find that both $F$ and $p$ obey an nearly linear relation with $\gamma$. When $\gamma$ increase to a significant value of $0.5g$, we can still get an entangled state with $F\approx0.89$ and $p\approx0.08$, which is acceptable in practical experiments.

From above three simulations we can conclude that first step of our scheme, i.e. entanglement generation, has fine robustness under certain level of experimental errors, which lays a basic foundation for future experiments.

\subsection{Photon redistribution}
As discussed above, a laser field can be used to control the relative phase of prepared states. Since the time period of manipulation can be made very short by enlarging the energy shift, we ignore other possible process that may happen during the manipulation. In order to demonstrate the results of simulation, we define the probability of observing two photons in the same detector $P_s$. By denoting the relative phase added by the laser field as $\phi=E\delta{t}$, we can get from Eq. (\ref{redistribution}) that
\begin{equation}
P_s(\phi)=\frac{1}{2}(1+cos\phi).
\end{equation}
In our simulation, for each chosen group of parameters, the process is repeated by $10^4$ times. We choose a waiting time for the second photon as $T_2=100T$ and compute the probability $P_s$ as $P_s=\frac{number\ of\ events\ that\ two\ photons\ go\ into\ the\ same\ detector}{total\ number\ of\ two\ photon\ detection}$.

\begin{figure}
\includegraphics[width=0.95\linewidth]{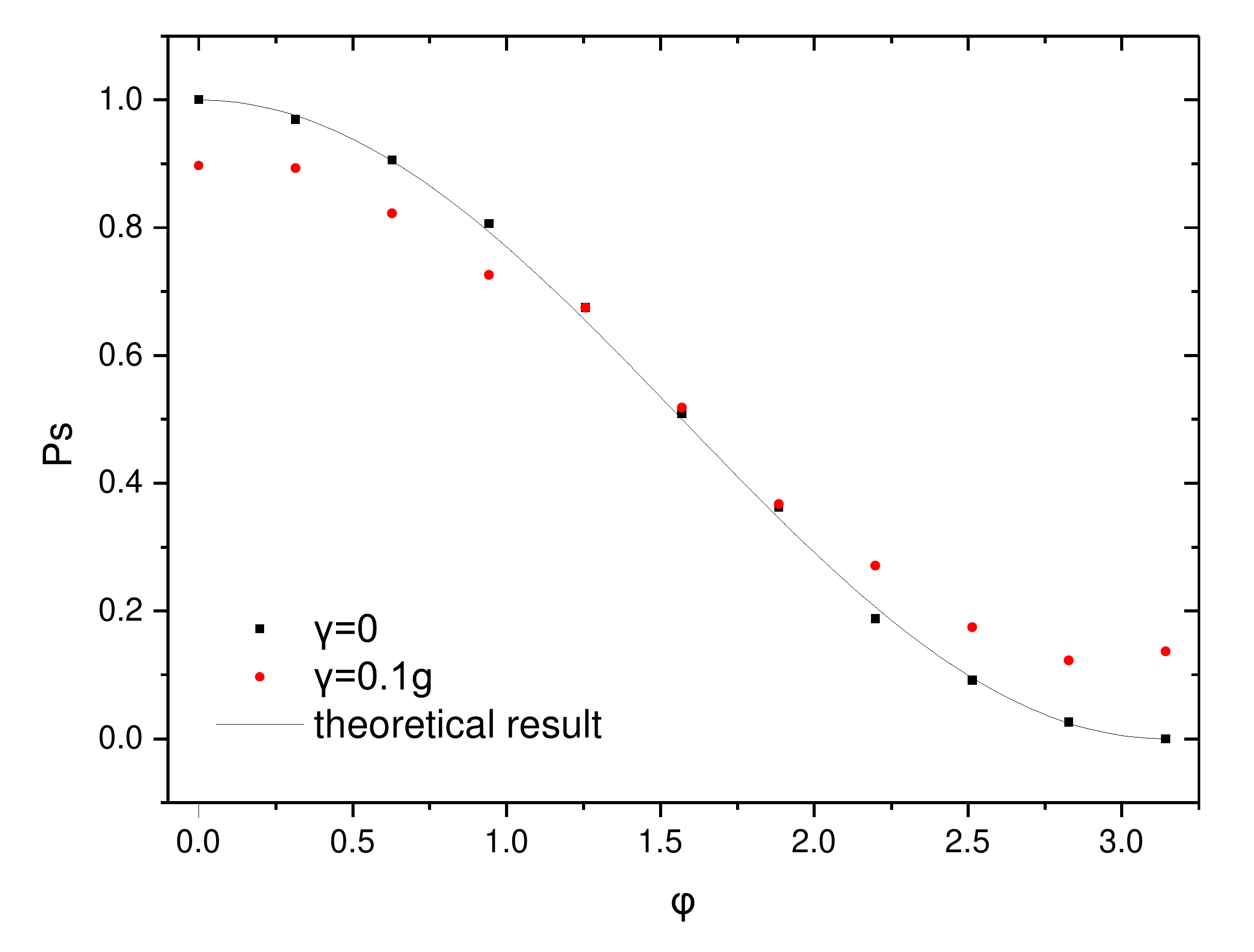}
\centering
\caption{Probability $P_s$ with varying levels of phase $\phi$. The black dots show results where spontaneous decay is neglect while red dots correspond to a spontaneous decay rate $\gamma=0.1g$. The black line shows the theoretical prediction $P_s(\phi)=(1+cos\phi)/2$.}
\label{f5}
\end{figure}

Here we consider two situations with and without spontaneous decay, i.e. evolution of the system is governed by $H_{eff}$ and $H'_{eff}$ respectively. The results of simulation are shown in Fig. \ref{f5}. According to the figure, a spontaneous decay rate $\gamma=0.1g$ gives $P_s$ certain deviation compared to theoretical results. However, when spontaneous decay rate is limited, in this case $\gamma=0$, the results fit very well with the theoretical curve. It can be conclude that even small spontaneous decay rate cannot be eliminated in practical experiments, the results can still obey the theory with significant precision.

In spite of that, we still have to point out that when spontaneous decay is included, the probability that the second photon detection happens in a certain waiting time will be relatively lower. For the case $\gamma=0$, within $T_2=100T$, one can basically ensure the detection of the second photon. When $\gamma=0.1g$, however, the probability is reduced to approximate $50\%$. This shows that $\gamma$ may have considerable influence on the success probability of our scheme.

\section{Discussion and conclusion}
In this paper, we demonstrate an approach of generating entanglement between two trapped ions in a HOM experimental setup. With the detection of a single photon, two ions are projected into a maximally entangled state. Using a quantum jump simulation, we take into account experimental errors including detector efficiency, beam splitter imperfection and the spontaneous decay of the ions. With rigorous analysis, we prove that this method has great robustness against all investigated errors. This makes it possible to realize the second step of our scheme. By manipulating the ions with a laser field (i.e. ac stark shift), we can modulate the evolution of the entangled state, which controls the detection of the secondly emitted photon. By changing the added phase, the second order correlation function of the two successively emitted photons can be arbitrarily controlled and the results of simulation fit well with our theoretical analysis. This idea gives us a new scenario of the HOM effect that it can be realized by temporally separated photons, where two-photon interference does not exist. Moreover, the controllable entanglement of the two ions may have potential applications in quantum information science.

\end{document}